\documentclass[aps,pra,reprint,groupedaddress]{revtex4-1}
\usepackage{graphicx}
\usepackage{color}
\graphicspath{{figure/}}
\usepackage{amsmath}
\usepackage{hyperref}
\graphicspath{{figure/}}
\usepackage{makecell}
\usepackage[all,defaultlines=1]{nowidow}

\begin{document}

\title{Silicon nitride based integrated photonic circuit to control a cold-atom source}

\author{H. Snijders$^1$, S. Hello$^2$, B. Wirtschafter$^1$, G. Feugnet$^1$, L. A. Tran$^3$, \\
M. M. Zafar$^3$, R. Dekker$^3$, C. I. Westbrook$^4$, A. Brignon$^1$, M. Dupont-Nivet$^1$}

\affiliation{${}^{1}$Thales Research and Technology France, 1 avenue Augustin Fresnel, 91767 Palaiseau, France \\
${}^{2}$Thales AVS France SAS, 40 rue de la Brelandière, 86100 Châtellerault, France \\
${}^{3}$LioniX International, Hengelosestraat 500, 7521 AN Enschede, Netherlands \\
${}^{4}$Laboratoire Charles Fabry, Institut d'Optique Graduate School, 2 avenue Augustin Fresnel, 91127 Palaiseau, France}
\date{\today}

\begin{abstract}
We have developed a silicon nitride based photonic integrated circuit (PIC) that is responsible for the cooling, pumping and imaging of cold rubidium 87 atoms.
The photonic integrated circuit consists of two chips placed next to each other and has a total area of 2x2~cm$^2$. 
This greatly minimizes the area needed while still having all the optical control functions to create, control and measure a magneto-optical trap (MOT). 
The piezo electric material Lead Zirconate Titanate (PZT) on the PIC is employed for phase shifting a Mach-Zehnder type configuration where extinction ratios up to 50 dB and switching speeds of 1 MHz are achieved. 
For the first time a two and three dimensional rubidium 87 MOT is realized using an active PIC. 
For the three-dimensional MOT, we measure $7\cdot 10^7$ atoms with a temperature of 270~$\mu$K. 
\end{abstract}

\maketitle


Cold atom interferometers have aroused great interest for the realization of inertial sensors due to their excellent stability and sensitivity \cite{Geiger2020}. 
Combining three accelerometers, three gyrometers and one clock, one can build an inertial measurement unit, which is an essential part of a navigation system, embedded in carriers such as aircraft, submarine, drone and satellite \cite{Barbour2010}. 
This device, installed on moving platforms, delivers information to computate inertial navigation solutions \cite{Jekeli2005,Savage1998a,Savage1998b}. 
One of the main challenges for cold atom inertial sensors is to significantly reduce their size to become useful for inertial navigation applications. 
This reduction can be achieved by combining two technologies: an on-chip atom interferometer \cite{Bohi2009,Schumm2005,DupontNivet2014, Reichel1999, Treutlein2004} and a compact optical laser system for cooling and control of the atoms.

To use cold rubidium atom technologies for atomic clocks in space \cite{Aveline2020,Laurent2020}, for on board gravimetry \cite{Bidel2018,Bidel2020}, or microgravity experiments \cite{Geiger2011,Muntinga2013,Rudolph2015,Becker2018} a lot of effort has been made to develop and ruggedize the laser system needed to cool and manipulate atoms. 
There are two main approaches to generate laser light at 780~nm. One can use laser diodes in extended cavities at 780~nm with semiconductor amplifiers \cite{Schmidt2011,Schmidt2011b, Cheinet2006,Merlet2014,Zhang2018} or use frequency doubling of telecom lasers at 1.56~$\mu$m \cite{Lienhart2007,Carraz2009, Theron2015,Theron2017, Diboune2017,Dingjan2006, Nyman2006, Stern2009, Menoret2011}. 
Compared to direct emission at 780~nm, this approach provides higher power \cite{Sane2012} and the whole system can be fully fiber based \cite{Leveque2014,Legg2017,Diboune2017}. 
Both systems need a free space optical system for routing and switching the laser beams needed in the device. 
People have developed miniaturized optical benches using miniature optics and special bonding technology with volumes of 4.4~L \cite{hello_miniaturized_2025} and 23 L \cite{zhu_miniaturized_2024} but these sizes could be massively reduced when using a Photonic Integrated Circuit (PIC). 
A PIC is a chip-based circuit that routes, controls and manipulates light on a chip.

\begin{figure}
	\centering  \includegraphics[width=0.48\textwidth]{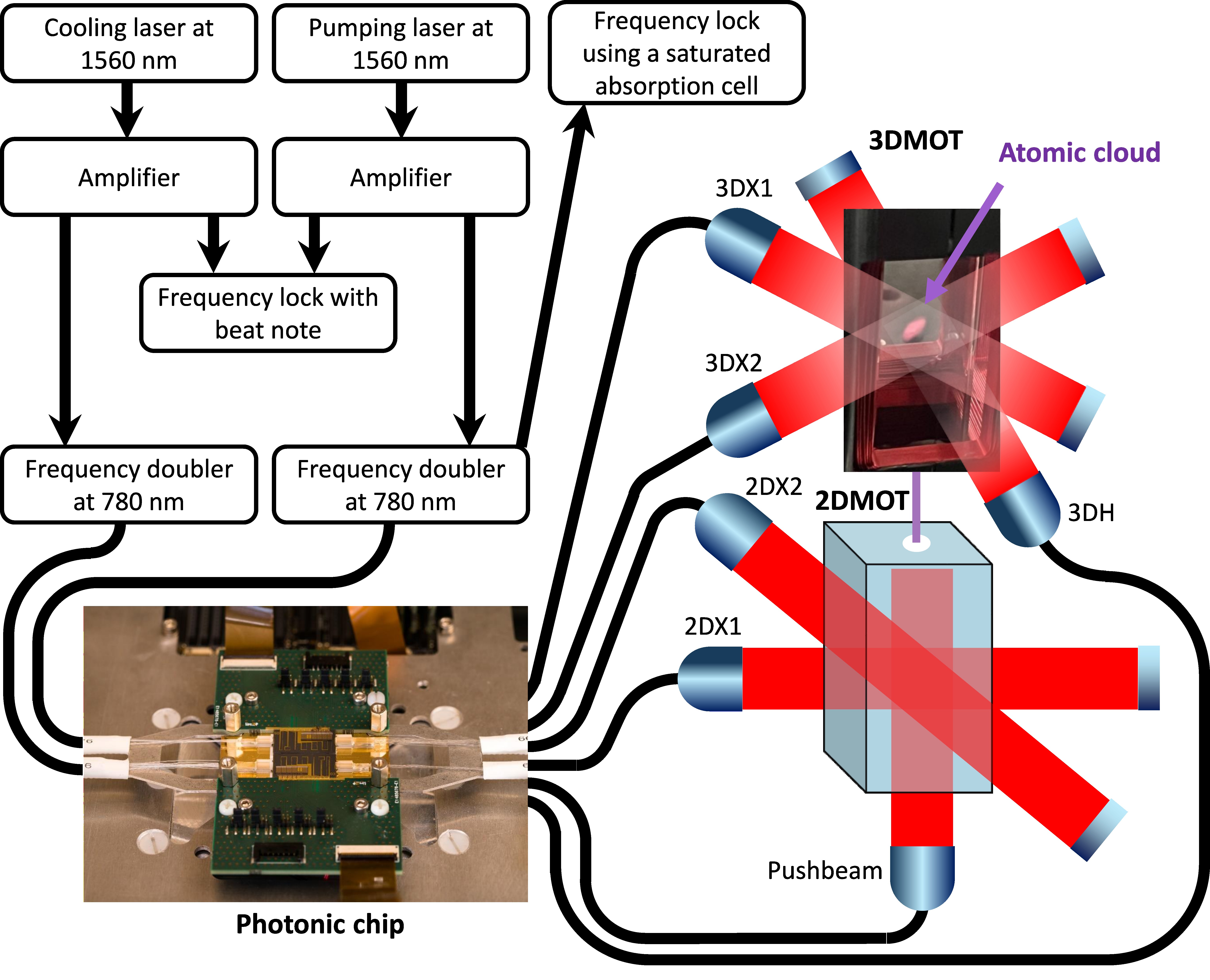}
	\caption{\label{fig_02}Layout of the experimental setup consisting of: a functional diagram of the fiber-based laser system for the cooler and pumper beams, a PIC for distribution of the laser light and the experimental setup for creating the 2D and 3D MOT.
	The photograph of the 3DMOT cell shows the atomic cloud in the center together with a sketch of the laser beams.}
\end{figure}

To realize an on-chip based inertial sensor using ultracold $^{87}$Rb atoms one needs to cool, trap and prepare the atoms in a specific state \cite{Huet2013,DupontNivet2016,Wirtschafter2022}. 
The first preparation step is Doppler cooling.
When combined with a magnetic field created using coils, this process creates a magneto-optical trap (MOT).
The laser system \cite{hello_miniaturized_2025} consists of two 1560 nm lasers, one for cooling and one for optical pumping, both are amplified and frequency doubled using a PPLN waveguide to reach 780~nm. 
In the supplemental material the level structure of $^{87}$Rb is shown to indicate the cooling and pumping transitions. 
2~mW of pumping laser power is sent to a saturated absorption spectroscopy setup (see Fig. \ref{fig_02}) to lock the pumping laser to the overlap between the $\left|5^2 S_{3/2},F=2\right>$ to $\left|5^2 S_{1/2},F=1\right>$ and $\left|5^2 S_{3/2},F=1\right>$ to $\left|5^2 S_{1/2},F=1\right>$ transitions since this line results in the highest intensity peak, afterwards an AOM is used to shift the pumper laser by +78.5~MHz. 
The cooler beam is locked to the pumper beam by sending 3~mW of power of both beams via a fibersplitter to a photodiode to generate a beatnote at 1560~nm. 
In Fig. \ref{fig_02} a schematic layout of the setup combined with a picture of the PIC and photograph of the atomic cloud is shown. 
After preparing the light in the laser setup, it is send to the PIC and its outputs are directly send to the fiber collimators of the MOT beams. 
The PIC replaces the bulk optics needed for control and manipulation of the laser beams. 
A 3DMOT is loaded in the upper cell from a 2DMOT realized in the lower cell (see Fig. \ref{fig_02}) where the atoms are pre-cooled and pushed using a Pushbeam into the 3DMOT \cite{Dieckmann1998,Schoser2002}.
Taking the three outputs 3DX1, 3DX2 and 3DXH from the PIC indicated in Fig. \ref{fig_02}, inserting them in the upper cell and reflecting them using mirrors creates six cooling beams with appropriate circular polarization. 
One pair of coils, whose axis is parallel to the 3DH beam, creates a quadrupole field with a gradient of 4~G.cm$^{-1}$ with a zero at the intersection of the six 3DMOT beams. 
The outputs 2DX1, 2DX2 and Pushbeam are combined with a magnetic gradient of permanent magnets for a 2DMOT. 

In this paper, we show the operation of a cold atom source employing a PIC that is used to distribute the laser beams to multiple MOT beams, control the power of each of the beams and enables switching these beams on a $\approx$~1~$\mu$s time scale by applying a phase shift to the waveguide using a piezo electric material (PZT). 

Piezoelectric actuation in photonic devices offers a low power, high speed solution and is compatible with materials such as Si and low loss Si$_{3}$N$_{4}$. 
These materials do not exhibit electro-optic effects since they have essentially no $\chi^{(2)}$ non-linearity due to their inversion symmetry \cite{tian_piezoelectric_2024}.
Piezoelectricity is an effect where charges can be generated and accumulated at the surface of a material, with no inversion symmetry, under external pressure. 
The inverse of this process can also happen by applying a current to induce stress and strain in a material. 
Stress is the internal resistance developed by the object to avoid deformation, while strain is the ratio of change in dimensions to the initial dimensions of the object. 
The stress and strain in the piezo electric material will cause stress to the waveguide material, leading to a change in the effective refractive index through the stress-optic effect. 
The key to induce this effect is to use a piezoelectric material with large piezoelectric coefficients. 
One material is PbZr$_{x}$Ti$_{1-x}$O$_{3}$ (PZT) with piezoelectric coefficients d$_{33f} >$~200~pm/V and e$_{31f} >$~18~C/m$^2$, which have been achieved using a wafer-scale Pulsed Laser Deposition growth method built by LAM Research \cite{nguyen_strongly_2017, nguyen_wafer-scale_2017}. 
To enhance the refractive index change in Si$_3$N$_4$ waveguides with PZT, LioniX International has developed a dome like structure of the PZT on top of the waveguide. 
This structure significantly increases the y-stress tensor present in Si$_3$N$_4$ \cite{hosseini_stress-optic_2015, everhardt_ultra-low_2022} and allows to obtain a $\pi$ phase shift by applying V$_{\pi}$~$\sim$~16~V at 1550~nm.      

\begin{figure}
	\centering  \includegraphics[width=0.50\textwidth]{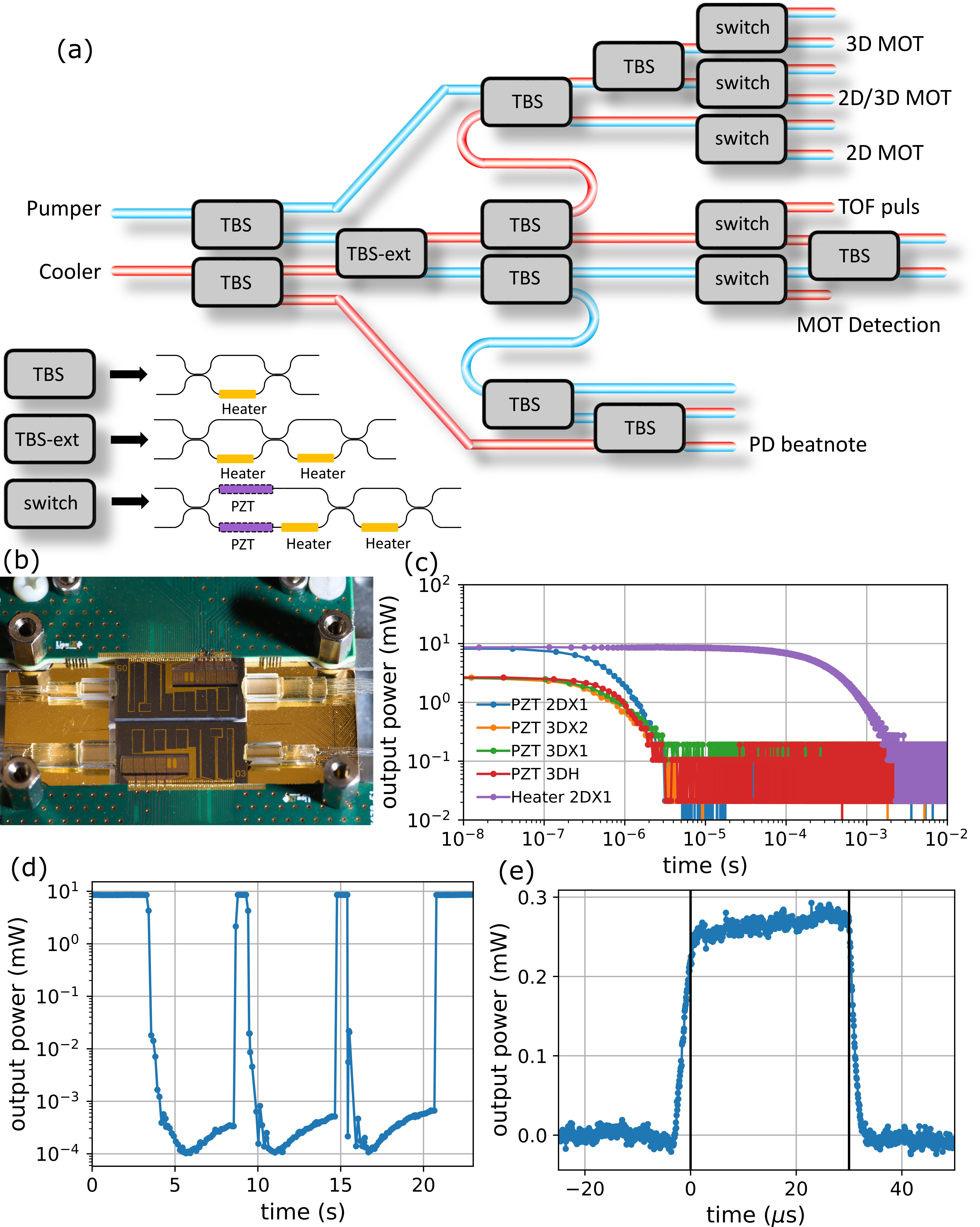}
	\caption{\label{fig_01}
		(a) Functional diagram of one of the photonic chips showing the paths of the cooler (red) and pumper (blue) laser beam in the circuit.
		(b) Photograph of the PIC consisting of two identical chips.  
		(c) Typical individual optical response of four PZT elements and one heater. 
		(d) Measured output power (with a S121C Thorlabs powerhead) as function of time while we switch on/off the voltage applied on the 2DX1 PZT element in order to switch on/off the light.
		The extinction ratio of 40-50~dB is limited by the stability of the PZT element.
		(e) 30~$\mu$s square pulse created using a PZT switch.}
\end{figure}

The PIC consists of two identically designed photonic chips and is shown in Figs. \ref{fig_01}.a and b. 
Each chip consists of a silicon base with on top a single stripe Si$_3$N$_4$ waveguide with a cross-section of 0.9~$\mu$m~$\times$~75~nm embedded between two layers of silicon oxide \cite{roeloffzen_low-loss_2018}. 
The waveguides are designed for single TE-mode operation at 780~nm. 
From the waveguides three main building blocks are designed, and used on the chip (see \ref{fig_01}.a): a tunable beam splitter (TBS), an extended TBS (TBS-ext) and a switch. 

The \textit{TBS} consists of two directional couplers with a heater between them which is on top of one of the waveguide arms and acts as a phase shifter. 
This TBS is equivalent to a Mach-Zehnder interferometer (MZI). 

The \textit{TBS-ext} is an extention of a TBS with an additional phase shifter and directional coupler leading to a new building block \cite{bandyopadhyay_hardware2021, suzuki_ultra-high-extinction-ratio_2015, wang_tolerant_2020}. This enables high extinction even in the presence of fabrication imperfections. Due to the high extinction ratio this building block also allows to geometrically cross two waveguides with respect to each other (see supplementary material).

The \textit{switch} is similar to the TBS-ext with the addition of a PZT actuator to both the top and bottom arm of the first phase loop of the interferometer (see Fig. \ref{fig_01}.b). 
A PZT actuator is constructed on both arms for additional redundancy. 
The advantage of a PZT actuator is that it allows MHz switching as opposed to kHz switching for a thermal actuator.

To obtain a MZI with one input and two outputs where it is possible to find a phase configuration such that the light is zero at one of the outputs, requires very specific value of the two directional couplers splitting ratio $k_i$, which is complicated due to fabrication errors. 
In a directional coupler, $k_i$ is the fraction of light that couples from one waveguide to another. 
Thus, another phase shifter and directional coupler (Fig. \ref{fig_01}.a) are added in TBS-ext and switches. We show in the supplementary material that this reduces the requirements on $k_i$ while still having zero transmission at one of the outputs for a specific phase configuration.
This allows us to make active waveguide crossings and high extinction switches on a PIC. 
Combining this with a piezo electric material like PZT enables the construction of fast switches with high extinction ratios and short detection pulses, which is important in our experiment. 
 
A functional layout of the photonic chip is shown in Fig. \ref{fig_01}.a which consists of 9 TBSs, 1 TBS-ext, and 5 switches. 
So that there are in total 10 PZT actuators and 21 thermal heaters on 1 chip. 
The entire assembly consists of two, up to fabrication imperfections, identical photonic chips to minimize the design complexity. 
Each chip has 2 optical inputs (cooler and pumper) and 13 outputs. 
The bottom 3 optical outputs can be used to generate a beatnote between the cooler and pumper beam and to send the pumper beam to a saturated absorption spectroscopy lock. 
The 4 center optical outputs are used as detection beams for cooler, pumper or both depending on the settings of the switch or for optical pumping. 
Finally, 3 of the top 6 outputs are used to cool the atoms in the 2DMOT and 3DMOT, while the other 3 outputs are used to remove unwanted light during switching and prevent on chip light dissipation. 
The chips are glued to a copper mount with one rotated by 180~degrees with respect to the other to attach the wirebonds. 
On both sides of the chip, fiber arrays are placed for coupling the light into and out of the chip. 
Mode overlap simulations, between the tapered waveguide and the fiber core, show an optimum of 0.7~dB of loss at 780~nm. 
Experimentally, we find a fiber-chip coupling loss of 1-2~dB. 
The photonic assembly is mounted on a peltier element for thermal stability. 

In Fig. \ref{fig_01}.c, the temporal response of the output power is plotted for 4 PZT elements with a turn off switching speed of 1-2~$\mu$s (mainly limited by the driving electronics) and for 1 heater with turn off speed of 1~ms. 
Turn off switching speed is here defined as time needed to achieve a $\pi$ phase shift after the element receives the command. 
The data is obtained with Thorlabs DET025AFC detector having a bandwidth of 2~GHz but a dynamic range limiting the extinction ratio measurements. 
Thus in Fig. \ref{fig_01}.d the extinction ratio of 1 of the PZT elements, measured with a thorlabs S121C powerhead, is plotted with a slower time scale. 
Here an extinction ratio of 40-50~dB is obtained, but is limited by drifts in the PZT due to the high accuracy needed to set the phase of the switch building block (see supplemental material). 
To switch the PZT actuator from 0 to $\pi$ a voltage between $V_{\pi} = 12-15$~V is needed. 
This architecture allows one to temporally shape the laser pulse, see Fig. \ref{fig_01}.e, where we created a 30~$\mu$s square pulse, used in the experiment as detection pulse. 

After frequency doubling, two 50/50 splitters are used to split both the cooler and pumper beams in two paths such that the light enters the two photonic chips. 
The power entering each photonic chip is 80~mW for the cooler and 13~mW for the pumper. 
Apart for the cell for saturated absorption spectroscopy the light in the entire setup up to the output of the PIC is in a fiber or in a PIC and thus removes any free-space alignment issues.  
The chip output powers are given in table \ref{tabel_01} for two cases: the optimal case and experimentally applied case. 
By design only 87.5\% of the cooler and 78.5\% of pumper beam input in the chip can go to the 2D and 3D MOT. 
The remaining light is used for the detection and the generation of a beatnote to lock the two lasers (a PIC function currently not used). 
The PIC is by design not fully configurable which minimizes the complexity and reduces the loss. 
Leading to an average transmission for the cooler (pumper) beam through both chips is 27\% (33\%).
The variation between the two is due to fiber-chip coupling losses which vary slightly from fiber to fiber. 
Experimentally the total transmission of all MOT beams is on average reduced by 30\% to realize the necessary extinction ratios of the MOT beams when switching (see column experimentally applied in table \ref{tabel_01}). 
This is needed because the design of the switch does not allow to switch between the global maximum and global minimum transmission using only one PZT actuator (see supplementary material). 
This adds loss when using this switching mechanism on a PIC.  

\begin{table}
	\resizebox{8.75cm}{!} 
	{
	\begin{tabular}{||c|c|c|c|c||} 
		\hline
		  & \multicolumn{2}{|c|}{optimal} & \multicolumn{2}{|c|}{ experimentally applied} \\
		\hline
		MOT beam & pumper [mW] & cooler [mW] & pumper [mW] & cooler [mW]  \\ [0.5ex] 
		\hline\hline
		2DX1 & 1.4 & 10.2 & 0.9 & 7.10  \\ 
		\hline
		2DX2 & 1.4 & 12.0 & 1.0 & 8.80  \\
		\hline
		Pushbeam & 1.0 & 3.6 & 0.5 & 2.00  \\
		\hline
		3DX1 & 1.0 & 3.7 & 0.8 & 2.9  \\
		\hline
		3DX2 & 1.0 & 3.7 & 0.7 & 2.9  \\ 
		\hline
		3DH & 1.0 & 4.0 & 0.8 & 3.1  \\ 
		\Xhline{3\arrayrulewidth}
		Total & 6.8 & 37 & 4.7 & 27 \\ 
		\hline
	\end{tabular}
	}
	\caption{\label{tabel_01} Optical power of the cooler and pumper beams needed to create the 2D and 3D MOT. 
	Two configurations are shown, one called 'optimal' for the highest power value obtained from the photonic chip and another configuration called 'experimentally applied' which has about 30$\%$ less power, to optimize the switch extinction ratio.}
\end{table}   

\begin{figure}
	\centering  \includegraphics[width=0.48\textwidth]{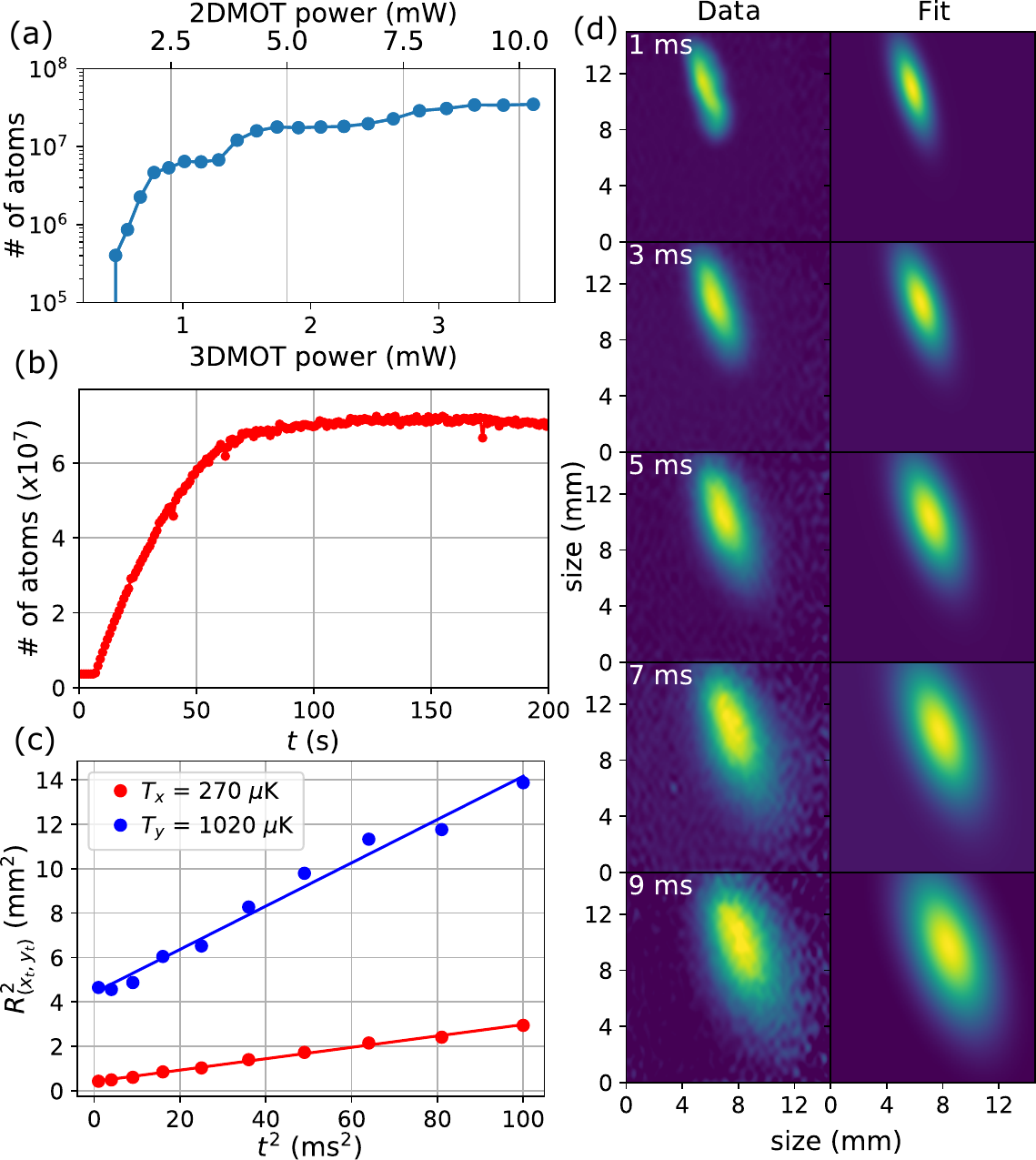}
	\caption{\label{fig_03}Experimental results for the atom number and temperature of the 3DMOT. 
	(a) atom number in the MOT as a function of the average cooler power (per beam) for the 2D and 3DMOT beams. For each measurement the loading time is 180~s, which is sufficiently long for reaching saturation of the trapped atom number.
	(b) Loading curve of the atom number in the 3DMOT as a function of time for an average 3DMOT cooler power (per beam) of 3.7~mW. 
	(c) TOF measurement indicating a 3DMOT temperature between 270~$\mu$K and 1~mK. 
	The lines are fits. 
	(d) 2D images of the expansion of the 3DMOT during the TOF with obtained optical density (left) and their Gaussian fit (right).}
\end{figure}

In Fig. \ref{fig_03}.a the atom number in the MOT as a function of the average cooler laser power for the 3DMOT and 2DMOT is plotted. 
The fluorescence of the MOT is measured using a photodiode \cite{Lewandowski2003}. 
The plot has at the beginning a steep increase in the number of atoms and flattens off for higher powers. 
Fig. \ref{fig_03}.b shows a typical fluorescence loading curve of the MOT for an average power, over the 3DMOT beam cooler powers, of 3.7~mW. 
After 120~s, we obtain a maximum of around $7\times 10^7$~atoms.

To obtain the temperature of the 3DMOT a time of flight (TOF) measurement was performed \cite{Wirtschafter2022, Brzozowski2002}. 
For the TOF all the MOT beams and the magnetic field are turned off such that the cloud of atoms can freely drop and expand. 
We first turn off the 2DMOT beams and Pushbeam, then increase the magnetic field linearly for 3 ms from $\sim$4 to $\sim$8~G.cm$^{-1}$. 
After the 3~ms stage of confining the MOT the 3DMOT laser beams and magnetic field are turned off at the same time. 
The TOF is defined as the time between turning off the 3DMOT beams and magnetic field and sending in the detection beam, with the same frequency as the cooler, a peak power of 0.26~mW and pulse duration of $\approx$~30~$\mu$s (Fig. \ref{fig_01}.e).

In absorption spectroscopy the atoms absorb the light of the detection beam such that the shape of the atomic cloud is visualized on the camera as a dark intensity region. 
Assuming the cloud of cold atoms is a Gaussian with radius $R_{(x_{0}, y_{0})}$ before expanding, due to the velocity distribution the cloud radius expand as: 
\begin{equation} 
R_{(x_{t}, y_{t})}^2 = R_{(x_{0}, y_{0})}^2 + \frac{k_{b}T}{M} \times t^2 \;,
\label{eq-TOF}
\end{equation} 
where $k_{b}$ is the Boltzmann constant, $M$ is the atomic mass, $T$ is the average temperature of the cloud and $R_{(x_{t}, y_{t})}$ is the measured radius of the cloud in the $x$ or $y$ direction and $t$ the TOF. 
The temperature is given by the slope of the graph $R_{(x_{t}, y_{t})}^2$ versus $t^2$. 
The result is shown in Fig. \ref{fig_03}.c, displaying a cloud temperature of 270~$\mu$K for the $x$-direction and 1~mK for the $y$-direction. 
For each TOF step, a 2D rotatable elliptical Gaussian fit was performed on the optical density (see supplementary material). 
Several images of the expanding cloud are shown in Fig. \ref{fig_03}.d as well as their corresponding fits. 
To create the atomic cloud with the lowest temperature the power of the 3DH cooler beam was reduced to 2~mW to balance the forces in the MOT since the force on the atoms due to the magnetic field is to first order twice as high in the 3DH direction as in the other perpendicular directions. 
The discrepancy between the two temperatures is most likely due to a slight imbalance of the MOT forces when turning off the magnetic field. 
The magnetic field cannot be turned off instantly and its corresponding force is not isotropic.  

In this paper, we demonstrate for the first time the operation of PZT switches responding on a $\approx$~1~$\mu$s time scale while also reaching extinction ratios up to 50 dB using a Si$_3$N$_4$ based PIC.
This unique PIC allows, to route, switch on and off, and temporally shape the output of the laser beams. 
The high extinction ratio allows us to switch the MOT beams during a TOF measurement without significantly disturbing the cloud.
It brings together all the optical functions for cooling, pumping and imaging in a volume of $2\times 2 \times 0.1$~cm$^3 \approx 4 \cdot10^{-4}$~L which is several orders of magnitude smaller compared to an optical bench of 4.4~L \cite{hello_miniaturized_2025} or a free space optical table of 600~L \cite{DupontNivet2025}. 
Operated with the PIC, the setup can load up to $7\times 10^7$ atoms in the 3DMOT at 270~$\mu$K in $\sim$100~s.
The atom number and loading rate are lower than what is obtained in \cite{hello_miniaturized_2025} ($2.5\times 10^8$) with 2 times more laser power for the 3DMOT and 5 times more for the 2DMOT. 
The obtained numbers can be enhanced by increasing the input power inside the PIC. 
Increasing the atom number up to $10^8$ is a prerequisite for an on atom-chip clock with a $10^{-13}$ relative stability \cite{DupontNivet2025, Szmuk2015} or an on atom-chip accelerometer with a 1~$\mu$g sensitivity \cite{DupontNivet2014}.
The demonstrated volume reduction of the optical system paves the way to a chip-based sensor for inertial navigation.
Further developments will be to integrate the beam collimators on a chip \cite{isichenko_photonic_2023,blumenthal_enabling_2024,huang_compact_2025,nshii_surface-patterned_2013} and emit light beams with a diameter of 1~cm from a grating coupler, which are positioned such that three beams create a MOT. 
Integrating such a system would provide an entirely on-chip and in-fiber integrated optical system for the atomic cloud control.

\vspace*{0.3cm}
\begin{acknowledgments}
Part of this work received funding from the European Defence Fund (EDF) under grant agreement 101103417 – project ADEQUADE. 
Funded by the European Union. Views and opinions expressed are however those of the author(s) only and do not necessarily reflect those of the European Union or the European Commission. 
Neither the European Union nor the granting authority can be held responsible for them.  
\end{acknowledgments}
 


\bibliography{biblio}

\widetext
\setcounter{equation}{0}
\setcounter{figure}{0}
\setcounter{table}{0}
\makeatletter
\renewcommand{\theequation}{S\arabic{equation}}
\renewcommand{\thefigure}{S\arabic{figure}}

\begin{center}
	\textbf{\Large Supplementary Materials}
\end{center}
	
\section{Extinction ratio of an extended Mach-Zehnder interferometer taking into account coupling errors}
	
\begin{figure}
	\centering  \includegraphics[width=1.0\textwidth]{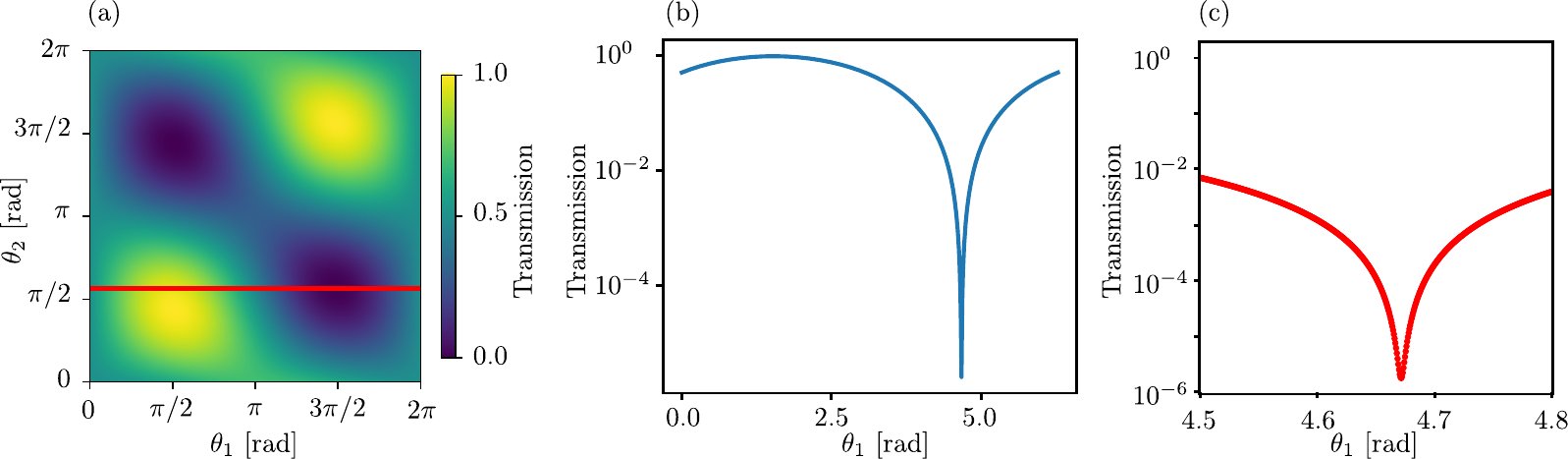}
	\caption{\label{sfig_2}(a) Simulation of the normalized transmission of an extended TBS as a function of $\theta_{1}$ and $\theta_{2}$ for the case where $k_{1} = 0.6$, $k_{2} = 0.4$ and $k_{3} = 0.5$. 
	In this simulation the light enters and exits from the top waveguide. 
	The horizontal line helps to visualize that the maximum and minimum are not aligned along the $\theta_{1}$ axis and one needs to change both $\theta_{1}$ and $\theta_{2}$ to switch between the global minimum and the global maximum. 
	(b) Transmission on a logscale as a function of $\theta_{1}$ for the cross section indicated by the red horizontal line in (a).
	(c) Zoom in around the section close to zero transmission of (b).}
\end{figure}

The advantage of an extended TBS (TBS-ext) is that high extinction ratios become possible since one is no longer limited by imperfections in the directional couplers for a TBS or Mach-Zehnder interferometer (Fig. \ref{fig_01}.a).
Compared to a standard TBS, a TBS-ext has an additional phase shifter region and directional coupler section. 
In this section we will explain why an TBS-ext helps us to become insensitive to fabrication errors of the directional couplers. 
To do this we start out by writing down the equation of a TBS using the matrix notation of the directional couplers and phase shifter. 
This leads to Eq. (\ref{TBS-equation}) which is written as
\begin{equation} 
	\begin{gathered}
		M_{TBS} =
		\begin{pmatrix}
			\sqrt{1-k_{2}} & -i\sqrt{k_{2}}\\
			-i\sqrt{k_{2}} & \sqrt{1-k_{2}}
		\end{pmatrix}
		\begin{pmatrix}
			1 & 0\\
			0 & e^{-i\theta_{1}}
		\end{pmatrix}
		\begin{pmatrix}
			\sqrt{1-k_{1}} & -i\sqrt{k_{1}}\\
			-i\sqrt{k_{1}} & \sqrt{1-k_{1}}
		\end{pmatrix}
	\end{gathered}
\label{TBS-equation}
\end{equation}
where $k_{1}$ and $k_{2}$ are the coupling values of the directional coupler describing the fraction of the light that couples from the top to the bottom waveguide and has a value between 0 and 1. 
$\theta_{1} = \left(2\pi \Delta n_{1} L_{1}\right)/\lambda$ is the phase shift applied to the bottom waveguide arm with $\lambda$ the wavelength of the light, $L_{1}$ the length of the waveguide arm and $\Delta n_{1}$ the materials change in refractive index due to an active element (here a thermal actuator or piezo-electric actuator). 
Assuming light enters the top input port of the TBS, written as the vector $\begin{pmatrix} 1 \\ 0 \end{pmatrix}$, the light coupled to the top output port of the TBS is
\begin{equation} 
\sqrt{(1-k_{1})(1-k_{2})} - \sqrt{k_{1}k_{2}}e^{-i\theta_{1}} \;,
\label{TBS-zero}
\end{equation}        
which will only be zero if $k_{1} + k_{2} = 1$ and if one also needs to switch over the full intensity range from $0$ to $1$ the condition becomes $k_{1} = k_{2} = 0.5$. 
In the presence of fabrication imperfection in $k_{1}$ and $k_{2}$, it is nearly impossible that these conditions are fulfilled, because having a small error $\Delta k$ on $k_{1}$ can to first order only be compensated for if one also has an error $-\Delta k$ on $k_{2}$ and the probability of exactly having two opposite errors that cancel each other is very small. 
Doing a similar calculation for the extended TBS one finds the following matrix equation	
\begin{equation} 
	\begin{aligned}
		M_{TBS ext} =
		\begin{pmatrix}
			\sqrt{1-k_{3}} & -i\sqrt{k_{3}}\\
			-i\sqrt{k_{3}} & \sqrt{1-k_{3}}
		\end{pmatrix}
		\begin{pmatrix}
			1 & 0\\
			0 & e^{-i\theta_{2}}
		\end{pmatrix}
		M_{TBS}
	\end{aligned}
\label{TBS-ext-equation}
\end{equation}
which leads under the same condition as Eq. (\ref{TBS-zero}) for the light coupled to the top output port
\begin{equation} 
\sqrt{1-k_{3}}\left(  \sqrt{(1-k_{1})(1-k_{2})} - \sqrt{k_{1}k_{2}}e^{-i\theta_{1}} \right)- \sqrt{k_{3}} e^{-i\theta_{2}} \left(  \sqrt{k_{2}(1-k_{1})} - \sqrt{k_{1}(1-k_{2})}e^{-i\theta_{1}} \right) \;,
\label{TBS-extension zero}
\end{equation}  
this equation clearly holds for a much larger range of $k_{1}$, $k_{2}$ and $k_{3}$ by adjusting the value of $\theta_{1}$ and $\theta_{2}$ to the proper values \cite{suzuki_ultra-high-extinction-ratio_2015, wang_tolerant_2020}. 
To estimate the necessary conditions (values of $k_{1}$, $k_{2}$ and $k_{3}$) for Eq. (\ref{TBS-extension zero}) to become zero, we first take the absolute value squared of Eq. (\ref{TBS-extension zero}) to remove the $\theta_{2}$ term. This gives equation
\begin{equation} 
\left|\sqrt{1-k_{3}}\left(  \sqrt{(1-k_{1})(1-k_{2})} - \sqrt{k_{1}k_{2}}e^{-i\theta_{1}} \right)\right|^2 = \left|\sqrt{k_{3}} \left(  \sqrt{k_{2}(1-k_{1})} - \sqrt{k_{1}(1-k_{2})}e^{-i\theta_{1}} \right)\right|^2 \;,
\label{TBS-extension abs}
\end{equation} 
which is then solved for $\theta_{1}$ and leads to
\begin{equation} 
\cos{\theta_{1}} = \frac{1 - k_{1} - k_{2} - k_{3} + 2k_{1}k_{2}}{2 \sqrt{k_{1}k_{1}(1-k_{1})(1-k_{2})}} \;.
\label{TBS-extension cos}
\end{equation} 
This equation is only valid if the right side of the equation has a value between $1$ and $-1$. 
The reason $\theta_{2}$ drops out when determining the extinction conditions is because of the symmetry in the equation and solution (see Fig. \ref{sfig_2}.a) and this means that once it is possible to find a solution $\theta_{1}$ there must also be a solution for $\theta_{2}$ such that extinction of the light is possible. 
Note however that for perfect extinction (zero transmission) one still needs to tune both $\theta_{1}$ and $\theta_{2}$ (see Fig. \ref{sfig_2}.a).     

We numerically evaluate the right side of Eq. (\ref{TBS-extension cos}) by calculating all combinations of $k_{1}$, $k_{2}$ and $k_{3}$ between $0$ and $1$ in steps of $0.01$. The results are plotted as a function of the effective radius from the ideal value $(k_{1} = 0.5, k_{2} = 0.5, k_{3} = 0.5)$ given by the equation  
\begin{equation} 
r_{eff} = \sqrt{(k_{1} - 0.5)^2 + (k_{2} - 0.5)^2 + (k_{3} - 0.5)^2} \;.
\label{TBS-extension reff} 
\end{equation} 
	
\begin{figure}
	\centering  \includegraphics[width=0.78\textwidth]{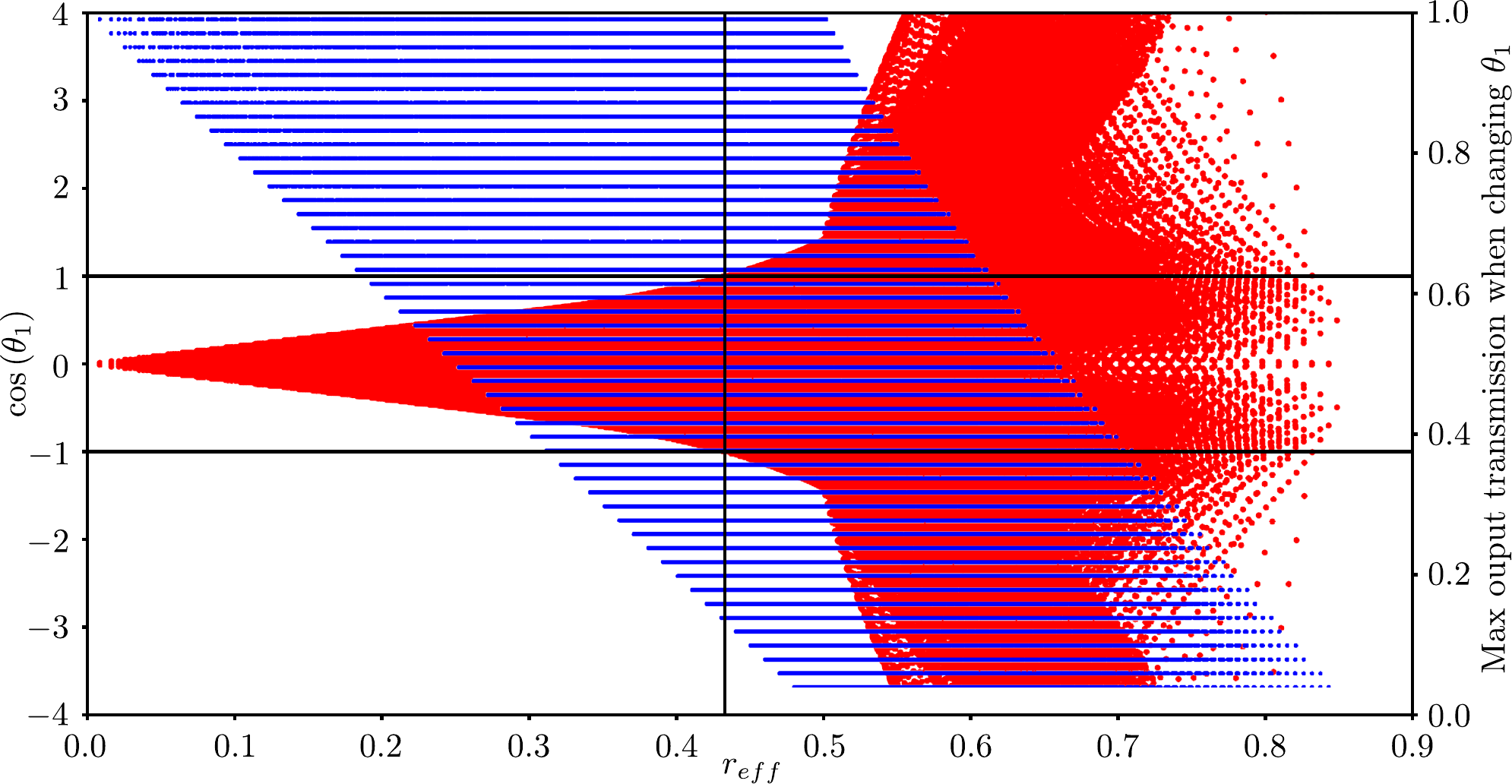}
	\caption{\label{sfig_1}Evaluation of Eq. (\ref{TBS-extension cos}) as a function of the effective radius $r_{eff}$. 
	The results are plotted as red dots. 
	The blue dots indicate the maximum output intensity that is reachable while keeping full extinction of the light when only changing $\theta_{1}$. 
	In this case the system acts as a single switch since only $\theta_{1}$ is changed to go from zero to the maximum output transmission, which is limited by the values of $k_{1}$, $k_{2}$, $k_{3}$.}
\end{figure}

In Fig. \ref{sfig_1} the right-hand side of Eq. (\ref{TBS-extension cos}) is plotted as function of $r_{eff}$. 
The two horizontal black lines indicate the $-1$ and $1$ values of $\cos{\theta_{1}}$. 
All values outside the region between the two horizontal lines indicate an area that is nonphysical since they do not represent a solution for Eq. (\ref{TBS-extension cos}). 
This marks the region in which there is always a solution to Eq. (\ref{TBS-extension cos}). 
Furthermore, considering the case where $k_{1}, k_{2}, k_{3} = 0.25$ or $0.75$ leads to the vertical black line at $r_{eff} = 0.433$. 
From this we deduce the conditions $(0.25 \leq k_{1} \leq 0.75), (0.25 \leq k_{2} \leq 0.75), (0.25 \leq k_{3} \leq 0.75)$ for which the light can always be extinguished in one of the outputs by finding the corresponding values for $\theta_{1}$ and $\theta_{2}$. 
There are bigger errors that can also be corrected but is dependent on the ratios of the coupling constants. Having fabrication errors inside the region $(0.25 \leq k \leq 0.75)$ is easily achievable for current photonic foundries at many different wavelengths. 
	
In the ideal case one would like to switch between the minimum and maximum value using only one active element (in this case PZT). 
Although there is relatively large range to find the zero-transmission value in the presence of fabrication errors it is in general not possible to also switch back to the global maximum (transmission value of 1) by only changing the $\theta_{1}$ phase. 
Note that it would be possible if both $\theta_{1}$ and $\theta_{2}$ are changed at the same time. 
The maximum output intensity that is achievable for each set of values given by $k_{1}$, $k_{2}$, $k_{3}$ is plotted with the blue dots in Fig. \ref{sfig_1}, where one observes that for a larger effective radius $r_{eff}$ the probability of switching back to the global maximum value using only $\theta_{1}$ decreases. 
The fact that the blue points are on horizontal lines is because many different $k$-values have the same absolute value for the output transmission, there complex phase differs however.        
	
This drawback of not reaching the maximum output intensity is illustrated in Fig. \ref{sfig_2}.a for the case of $k_{1} = 0.6$, $k_{2} = 0.4$, $k_{3} = 0.5$, where an optimum value of 0.95 is obtained when having a fixed $\theta_{2}$ and only changing $\theta_{1}$. 
But allowing to change $\theta_{2}$ again makes sure that the global maximum is attainable. 
Furthermore, Fig. \ref{sfig_2}.b and c illustrate that to reach high extinction ratios it is necessary to very accurately set the phase of the respective actuator and this is currently the limiting factor to achieve higher extinction ratios.

\section{comparison PZT and thermal actuators}
	
In this section, we give a comparison between PZT and thermal actuators.
Both are actuators used to change the phase in one of the waveguide arms of a Mach-Zehnder interferometer, however functionally there is an important difference between them.
The thermal heaters are used to steer a certain amount of the light beam to a specific output. 
The PZT on the other hand is used to quickly turn several laser beams on and off.
Table \ref*{tabel_02} shows a comparison between the performance of a PZT and thermal actuator in TBS or TBS-ext like configuration.

\begin{table}
	\resizebox{18cm}{!}
		{ 
		\begin{tabular}{||p{4cm}|p{7cm}|p{7cm}||} 
			\hline
			   & PZT actuator & Thermal actuator  \\ [0.5ex] 
			\hline\hline
			$V_{\pi}$ & 14~$\pm$~1~V & 7~$\pm$~0.5~V   \\ 
			\hline
			Switching time & 2-3~$\mu$s & 2-3~ms  \\
			\hline
			Actual actuator length & 1~cm (folded to minimize space) & 1~mm \\
			\hline
			Stability & Typically, the PZT can be set up to $\pm$~0.025 radian accuracy for the short term. A stability of 0.01 radian/hour can be reached but requires waiting up to a minute for the material stress to stabilize.  & Thermal stability is required, but once system is thermally stable, the drift of the induced phase-shift is below 0.01 radian/hour. This stability is achieved thanks to the equality of the two interferometer arms length.\\
			\hline
			Reproducibility & As it can be seen in Fig. \ref{fig_01}.d, the reproducibility, for short term operations, is good. For long term, we needed to slightly recalibrate the PZT every 2-3 weeks for proper operation with the control electronics. After the recalibration, the maximum phase shift drift we observed was 0.1 radians. This drift is also due to the drift of the PZT control electronics. & Once the heaters had been calibrated, we could run several months of experiments without recalibrating the system. The optical output power of each channel did not change by more than 1\%. This 1\% is currently the limit of the power stability of our laser system.   \\ 
			\hline
			Hysteresis & In our experiment, we wait after each switch several seconds up to a few minutes. On these timescales, we do not observe hysteresis. However, we have observed that when switching every second or faster several times, hysteresis effects become visible. & No hysteresis observed. \\ 
			\hline
		\end{tabular}
		}
	\caption{\label{tabel_02} Comparison between the performance of PZT and thermal actuators on the photonic integrated circuit reported in this work. }
\end{table} 

In Fig. \ref{sfig_2}.c, a zoom-in of Fig. \ref{sfig_2}.b is shown, which relates the phase shift to the obtained extinction ratio. 
To obtain a high extinction ratio an accurate setting of the phase is needed. 
This plot allows us to infer from the measured extinction ratio stability what phase stability has approximately been reached.

The chip is thermally stabilized with a peltier and kept stable within 0.1~K error margin given by the sensitivity of the temperature sensor. 
Here the temperature is measured in a copper part on which the PIC is glued. 
When changing the phase in a waveguide using a thermal actuator, no measurable temperature change is observed with the temperature sensor. 
Once all the thermal heaters are off and we turn them all on, then, this typically leads to a 0.5~K increase in temperature from the standard operation temperature of 300~K, but, within a minute, it is corrected by the peltier. 
After setting all the heaters to the optimum values, we wait a few minutes until a stable thermal equilibrium is reached, after that only the PZT is used to switch the light on and off. 
We have not observed thermal or other interference between the operation of the PZT and thermal actuators. 

\section{Hyperfine structure of rubidium 87}
	
\begin{figure}[h]
	\centering  \includegraphics[width=0.48\textwidth]{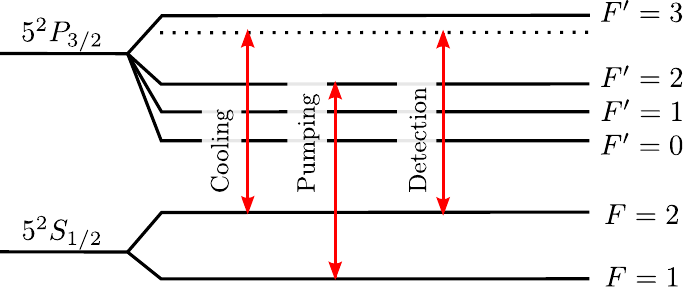}
	\caption{\label{sfig_3}Schematic drawing of the hyperfine structure of rubidium 87 and the laser frequency used for the cooling, pumping and detection beams.}
\end{figure}

In Fig. \ref{sfig_3} a schematic sketch of the hyperfine structure of rubidium 87 is given. 
The frequency used for the cooling corresponds to the transition $\left|5^2 S_{1/2},F=2\right> \rightarrow \left|5^2P_{3/2},F=3\right>$ of the hyperfine structure of rubidium 87 \cite{hello_miniaturized_2025,Steck2003} and works as if it is a closed cycle transition. 
However, this one is not fully closed because the frequency difference between $\left|5^2P_{3/2},F=3\right>$ and $\left|5^2P_{3/2},F=2\right>$ is only about 266~MHz, thus some atoms can go to $\left|5^2P_{3/2},F=2\right>$ leading to a possible decay in $\left|5^2 S_{1/2},F=1\right>$. 
After many absorption-emission cycles, all atoms will fall into that state and there will be no longer atoms in the cooling cycle. 
Therefore, a second laser (pumping laser) is needed to recycle atoms fallen into the $\left|5^2 S_{1/2},F=1\right>$ state back into the cooling cycle. 
The pumping laser will be resonant with the $\left|5^2 S_{1/2},F=1\right> \rightarrow \left|5^2 P_{3/2},F=2\right>$ transition. 
	
\section{Analysis of the camera images}
	
For the TOF the detection beam is injected perpendicular to the 3DH beam in the upper cell and detected with a Zelux camera. 
After obtaining the data image we apply a Fourier filter to remove high frequency interference fringes which appear due to the glass cell and alignment. 
Then we transform the absorption spectroscopy data to an optical density by using the following equation
\begin{equation} 
I(x,y) = I_{0}(x,y)e^{-D(x,y)} \;,
\label{eq-OD}
\end{equation} 
where $I(x,y)$ corresponds to the intensity of each pixel $(x,y)$ on the camera image with the atomic cloud present and $I_{0}(x,y)$ to the intensity of each pixel on the camera image without any atomic cloud present. 
Rewriting this equation leads to the optical density given by $D(x,y) = -\ln \left( \frac{I(x,y)}{I_{0}} \right) $. 
Each measurement destroys the atomic cloud, hence we repeat the measurement several times and every time measure the light absorbed by the atomic cloud at a different moment during the time of flight. 
To do this the camera is triggered with a hardware trigger. The measurement time in our case is not determined by the camera exposure time, but by the duration of the optical detection pulse. 
After each image a new background image without the cloud present is taken. 
The magnification is calibrated by visualizing measuring tape on the camera with an error below 1$\%$. 
Each TOF image is fitted with the equation 
\begin{equation} 
A + B  \exp{\left(-\frac{1}{2} \left( \frac{(x-x_{0}) \cos{\theta} - (y-y_{0}) \sin{\theta}} {R_{x_{t}}}  \right)^2 -\frac{1}{2} \left( \frac{(x-x_{0}) \sin{\theta} + (y-y_{0}) \cos{\theta}} {R_{y_{t}}}  \right)^2 \right) } \;,
\label{eq-gaussian fit}
\end{equation} 
which has seven free parameters. 
We define $x_{0}$ and $y_{0}$ as the center position of the atomic cloud, $R_{x_{t}}$ and $R_{y_{t}}$ are the standard deviations of the cloud along the two axis of the ellipse here named $x$ and $y$ and $\theta$ is the angle with which the cloud is rotated between the vertical direction and $y$-axis of the ellipse. 
The parameters $R_{x_{t}}$ and $R_{y_{t}}$ are used here to quantify the expansion of the cloud (see main text).  
	
\section{Calculation of atom number from the fluorescence response}
	
The number of atoms in the MOT can be determined by collecting part of the MOTs fluorescence with a photodiode \cite{Lewandowski2003}. 
The formula to calculate the number of atoms in the MOT is then given by the ratio 
	
\begin{equation} 
N_{atoms} = \frac{I_{MOT}}{R_{p}}
\label{number of atoms}
\end{equation} 
where $I_{MOT}$ represents the number of photons persecond emitted by the MOT and collected by the photodiode and $R_{p}$ is the number of photons scattered per atom. 
The number of photons scattered per atom is given by
\begin{equation} 
R_{p} = \frac{\Gamma}{2} \frac{\frac{I_{p}}{I_{sat}}}{1+\frac{I_{p}}{I_{sat}} + 4 \left( \frac{\delta}{\Gamma} \right)^2 } \;,
\label{photons scattered per atom}
\end{equation}  
where $\Gamma = 2\pi \times 6 \cdot 10^6$~Hz, $I_{sat} = 1.67 $~mW/cm$^2$ \cite{Steck2003} and we assume zero detuning ($\delta = 0 $). 
The incident intensity for all beams of the 3DMOT is $I_{p} = 1.7$~mW/cm$^2$. 
The formula for the number of photons emitted by the MOT and collected by the photodiode $I_{MOT}$ is given as
\begin{equation} 
I_{MOT} = \frac{S_{PhD}}{A_{gain} B_{response}} \frac{1}{h\nu} \frac{1}{T_{opt}} \frac{4\pi}{\Omega} \;,
\label{photons emitted by the MOT}
\end{equation} 
where $S_{PhD}$ is the signal of the photodiode in units of volt, $A_{gain}$ is the gain of the photodiode in units of $V/A$, and $B_{response}$ is the photodiode response in units of $A/W$, $h\nu = 2.54 \times 10^{-19}$~J is the energy of a single photon at 780~nm, $T_{opt}$ is the transmission of the glass cell and lenses before reaching the detector and $\Omega$ is the solid-angle of detection. 
	
\end{document}